\documentclass[aps,twocolumn,showpacs,superscriptaddress]{revtex4}
\usepackage{amsmath}
\usepackage{graphicx}
\usepackage{esint}
\usepackage{epstopdf}
\graphicspath{{pict/}{./}}
\usepackage{bm}%

\newcounter{Fig}

\newcommand{\be}{\begin{equation}}
\newcommand{\ee}{\end{equation}}

\begin{document}

\title{Scattering invisibility with free-space field enhancement of all-dielectric nanoparticles}
\author{Wei Liu}
\email{wei.liu.pku@gmail.com}
\affiliation{College of Optoelectronic Science and Engineering, National University of Defense
Technology, Changsha, Hunan 410073, P. R. China}
\author{Andrey E. Miroshnichenko}
\affiliation{Nonlinear Physics Centre,  Research
School of Physics and Engineering, Australian National University,
Canberra, ACT 0200, Australia}
\pacs{
        78.67.-n,   
        42.25.Fx,   
        78.67.Pt   
}

\begin{abstract}
Simultaneous scattering invisibility and free-space field enhancement have been achieved based  on multipolar interferences among all-dielectric nanoparticles. The scattering properties of all-dielectric nanowire quadrumers are investigated and two sorts of scattering invisibilities have been identified: the trivial invisibility  where the individual nanowires are not effectively excited; and the nontrivial invisibility with strong multipolar excitations within each nanowire, which results in free-space field enhancement outside the particles. It is revealed that such nontrivial invisibility originates from not only the simultaneous excitations of both electric and magnetic resonances, but also their significant magnetoelectric cross-interactions. We further show that the invisibility obtained is both polarization and direction selective, which can probably play a significant role in various applications including non-invasive detection, sensing, and non-disturbing medical diagnosis with high sensitivity and precision.
\end{abstract}
\maketitle

\section{Introduction}
\label{sec:intro}

The seminal topic of light scattering by nanoparticles (including both individual and clustered particles) has gained a new impetus from the recent rapid development of a nanophotonic branch based on resonant all-dielectric high-index nanostructures~\cite{Zhao2009_materialtoday,Liu2014_CPB,jahani_alldielectric_2016,KUZNETSOV_Science_optically_2016,staude_metamaterial-inspired_2017}. The co-excitation of electric and magnetic resonances together with their interference effects offers a new dimension of freedom for both near-field control and far-field scattering pattern shaping~\cite{Liu2014_CPB,jahani_alldielectric_2016,KUZNETSOV_Science_optically_2016,staude_metamaterial-inspired_2017,SMIRNOVA_Optica_multipolar_2016,LIU_ArXivPrepr.ArXiv160901099_multipolar_2016}.  Such extra freedom  further accelerates the development of other related nanophotonic branches of metasurfaces and metadevices, making accessible many functionalities and practical applications based on strong light matter interactions within low-loss resonant nanostructures~\cite{jahani_alldielectric_2016,KUZNETSOV_Science_optically_2016,CHEN_Rep.Prog.Phys._review_2016}, and especially make a critical step forward towards the future integration of nanophotonic and electronic chips and circuits.

\begin{figure*}
\centerline{\includegraphics[width=16.5cm]{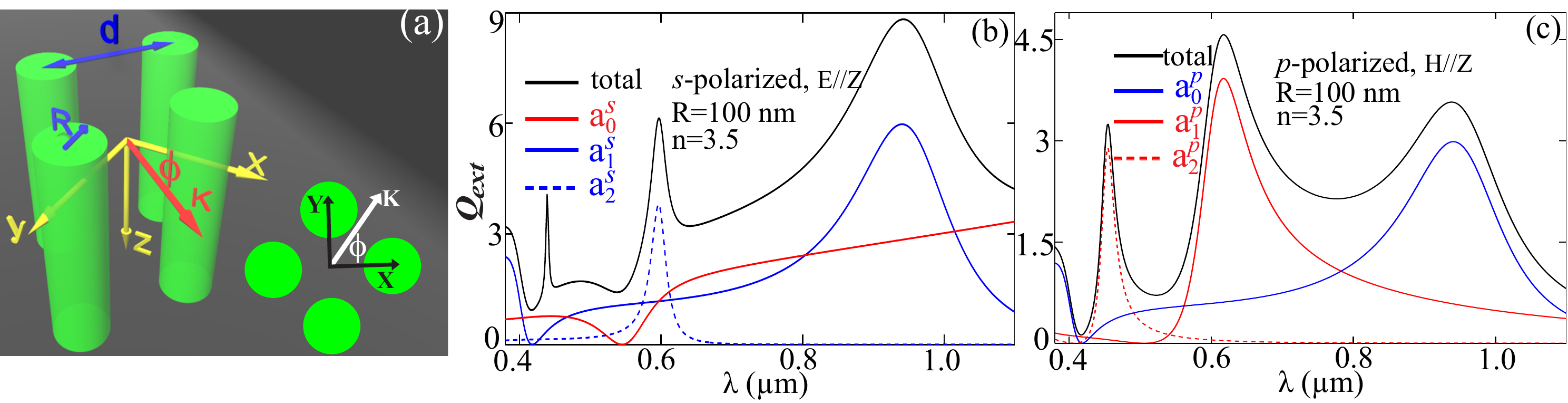}}\caption{ (a) Schematic of the scattering of a normally incident plane wave (with wave-vector $\mathbf{k}$ and polar angle $\phi$) by a symmetric quadrumer consisting of four identical parallel nanowires of refractive index $n$, radius $R$ and inter-particle distance $d$. The wave can be $s$- or $p$-polarized, which corresponds to electric or magnetic fields orienting along $z$ direction, respectively. (b) and (c) show the extinction efficiency spectra for a single nanowire of $n=3.5$ and $R=100$~nm for $s$- and $p$-polarizations respectively. Here both the total extinction efficiency spectra (black curves) and those spectra contributed by different individual resonances (blue curves: magnetic multipoles; red curves: electric multipoles) are shown.} 
\label{fig1}
\end{figure*}

Stimulated by the field of metamaterials and especially its flagship achievement of electromagnetic cloaking~\cite{SCHURIG_Science_metamaterial_2006,FLEURY_Phys.Rev.Appl._invisibility_2015}, recently the topic of scattering invisibility and transparency of nanoparticles has attracted surging interest (see Ref.~\cite{Alu2009_PRL,FLEURY_Phys.Rev.Appl._invisibility_2015} and references therein). Conventional approaches to render scattering particles invisible to a large extent rely either on guiding the impinging waves to flow around to avoid direct electromagnetic interactions or on the scattering cancellations of electric multipoles of opposite orientations~\cite{Kerker1975_JOSA,FLEURY_Phys.Rev.Appl._invisibility_2015}. Recently it has also been demonstrated that the complete destructive interferences of electric and toroidal multipoles can be employed to make scattering particles practically transparent ~\cite{Basharin2014_arXiv,miroshnichenko2014seeing,liu_toroidal_2015,Liu2015_OL_invisible,PAPASIMAKIS_NatMater_electromagnetic_2016}.
Nevertheless, for the traditional approaches mentioned above, usually the fields are only enhanced in enclosed isolated areas (inside the scattering particles for example), which are more or less inaccessible and as a result hinders many further applications.

For some specific applications such as non-invasive detection, sensing and even medical diagnosis without disturbance, to guarantee the high sensitivity and precision, usually invisibility together with simultaneous more accessible free-space field enhancement is required. This is of great significant for the enhancement of the interactions with various sensing and measuring devices, and at the same time can render enormous freedom for further field manipulations and signal processing, which is of course much more flexible in free space. It's well known that significant field enhancement can be widely found within various plasmonic nanostructures, which however are inevitably visible due to the absorption of metal.  Though it is shown that in all-dielectric particle clusters, Fano resonances that originate from hybridizations of electric and magnetic dipoles can be employed to significantly reduce the scattering cross sections~\cite{Miroshnichenko2012_NL6459,HOPKINS_ACSPhotonics_interplay_2015,YAN_ACSNano_directional_2015}, those scattering ensembles are still effectively visible. Recently it is demonstrated that such invisibility with free-space field enhancement is obtainable within all-dielectric metasurfaces through introducing symmetric gaps~\cite{YANG_NatCommun_alldielectric_2014}. Nevertheless, we should bear in mind that rendering large-scale extended metasurfaces or nanoscale scattering particle systems invisible are fundamentally different: for the former case, only the elimination of the reflection is needed when other diffractions are fully suppressed through sufficiently decreasing the periodicity; while for the later case, the effective scattering elimination along all directions are required. Basically it is of great importance while at the same time also intrinsically challenging to obtain simultaneous invisibility and free-space field enhancement with subwavelength scattering particle systems.

In this work, we study the scattering properties of all-dielectric quadrumers made of identical high-index nanowires. Besides the trivial invisibility obtained at the spectral positions where none multipoles of each individual nanowire are effectively excited and thus exhibit no field enhancement, we achieve also the nontrivial invisibility accompanied by free-space field enhancement within the gaps between the nanowires. Such nontrivial invisibility is revealed to originate from the significant co-excitations and magnetoelectric cross-interactions of all electric and magnetic multipoles, the functioning mechanism of which is contrastingly different from the conventional electric dipolar cancellations or electromagnetic anapole excitations~\cite{Kerker1975_JOSA,FLEURY_Phys.Rev.Appl._invisibility_2015,Miroshnichenko2012_NL6459}.  We further show that such nontrivial invisibility is dependent on not only inter-particle distances, but also both the polarizations and the propagation directions of the incident plane waves. It is expected that the mechanism we discover to obtain nontrivial invisibility relying on magnetoelectric cross-interactions between electric and magnetic multipoles can be generally extended to scattering clusters of other distributions (including periodic and random ones) with particles of other shapes. It may play a significant role in not only applications of non-invasive detections and sensing, but also in various scattering related fundamental researches, such as random lasers~\cite{WIERSMA_Nat.Phys._physics_2008}, $\mathcal{P}\mathcal{T}$-symmetric electromagnetic scattering~\cite{CHONG_Phys.Rev.Lett._mathcalpmathcaltsymmetry_2011} and topological photonics~\cite{Lu2014_topological,SLOBOZHANYUK_NatPhoton_threedimensional_2016}.

\section{Theoretical analysis based on multiple scattering method}
Figure~\ref{fig1}(a) shows schematically the scattering configuration we investigate: the quadrumer consists of four identical parallel dielectric nanowires of radius $R$, refractive index $n$, closest inter-particle (inter-surface) distance $d$ ($d-2R$); the coordinate positions of the nanowire centers on the $x-y$ plane are $(x,y)=(\pm d/\sqrt{2},0),(0,\pm d/\sqrt{2})$; the normally incident plane wave can be $s$-polarized (with electric field along $z$ direction, $\mathbf{E_0}\parallel\mathbf{z}$) or $p$-polarized (with magnetic field along $z$ direction, $\mathbf{H_0}\parallel~\mathbf{z}$), and its wave-vector $\mathbf{k}$  makes an angle of $\phi$ with respect to the $x$ axis. Throughout this paper, without losing generality, we set $n=3.5$ and $R=100$~nm. For the simplest case of single nanowire scattering of a normally incident plane wave, the extinction efficiency (extinction cross section divided by the diameter of the nanowire) can be expressed as~\cite{Bohren1983_book}:
\begin{equation}
\label{Q_ext_single}
Q_{\rm ext}^{s,p} = {2\over {kR}}\sum\limits_{m = -\infty}^\infty \mathbf{Re}(a_m^{s,p}),
\end{equation}
where $\mathbf{Re}(\cdot)$ means to take the real part, $a_m^{s,p}$ are the scattering coefficients for $s$- and $p$- polarizations, $m$ is the cylindrical harmonic number, and $k$ is the angular wavenumber in the background (it is vacuum in this study).  The rotational symmetry of a single scattering nanowire requires that $a_m^{s,p}=a_{-m}^{s,p}$ and thus the extinction efficiency can be further simplified as $Q_{\rm ext}^{s,p} = {2\over {kR}}[\mathbf{Re}(a_0^{s,p})+2\sum\nolimits_{m = 1}^\infty \mathbf{Re}(a_m^{s,p})]$. It is worth mentioning that~\cite{Zhao2009_materialtoday,kallos2012resonance,Liu2013_OL2621}: for $s$-polarization, $a_0^{s}$ corresponds to the electric dipole (ED), and $a_{1,2}^{s}$  correspond to the magnetic dipole (MD) and magnetic quadrupole (MQ), respectively; while for $p$-polarization, $a_0^{p}$ corresponds to the MD, and $a_{1,2}^{p}$ correspond to the ED and electric quadrupole (EQ), respectively. Moreover, it is easy to prove that $a_1^{s}=a_0^{p}$, which indicates that for both polarizations the MD resonances are exactly the same except that the magnitude for $s$-polarization is twice that for the $p$-polarization. This is also clear in Figs.~\ref{fig1}(b) and (c), where we show the extinction efficiency spectra (including both total extinction and those individual multipolar contributions) for both polarizations of a single nanowire.

\begin{figure*}
\centerline{\includegraphics[width=16.5cm]{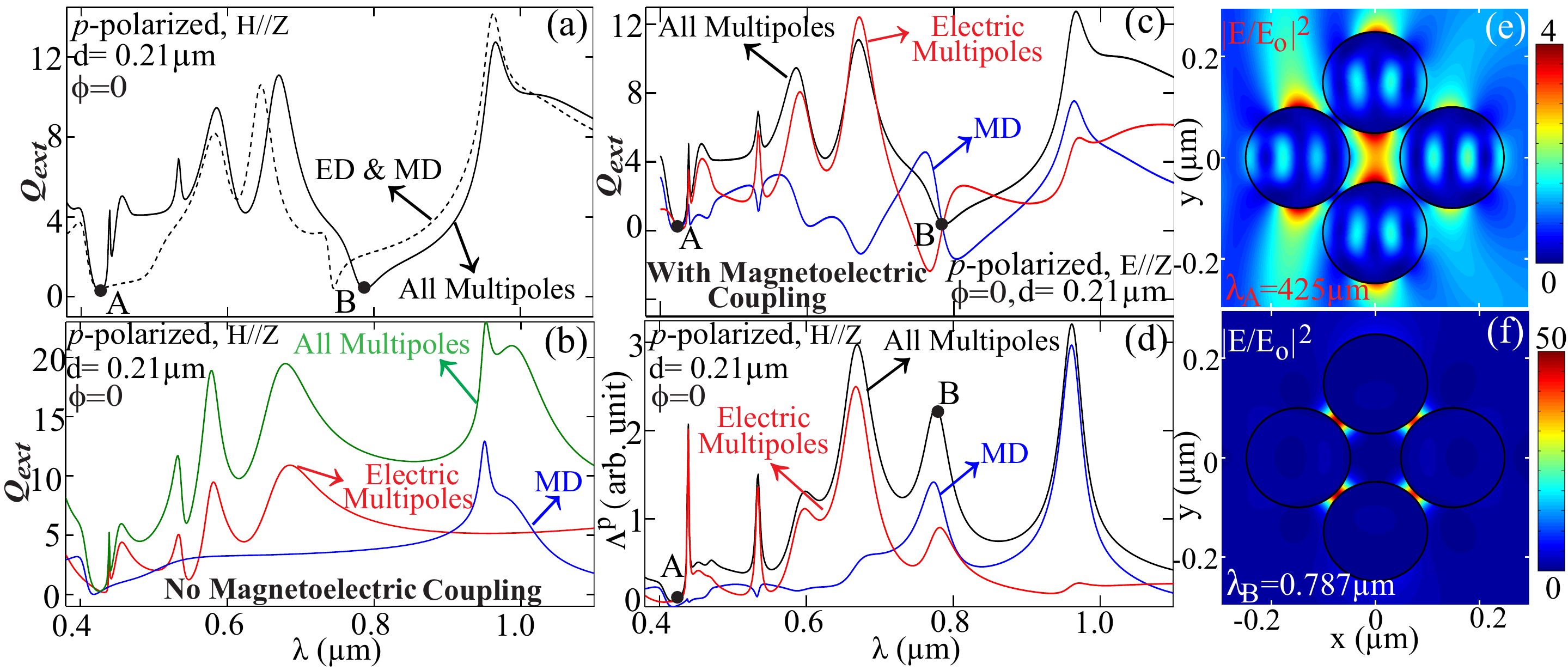}}\caption{Extinction efficiency spectra for the quadrumer of $d=210$~nm and $\phi=0$ with $p$-polarized incident waves: (a) Solid curve: full multipolar calculation; dashed curve: dipole approximation considering only the ED and MD. Two points where the quadrumer is effectively invisible are pinpointed by points $\mathbf{ A }$ ($\lambda_A=425$~nm) and $\mathbf{B }$ ($\lambda_B=787$~nm); (b) Extinction spectra obtained with the magnetoelectric coupling effects neglected, which are contributed by  MDs (blue curve),  electric multipoles (blue curve), and all multipoles (green curve), respectively; (c) Extinction spectra contributed by MDs (blue curve),  electric multipoles (blue curve), and all multipoles (black curve), respectively when the magnetoelectric coupling effects are considered. (d) Sum of scattering coefficient intensity: MDs (blue curve),  electric multipoles (blue curve), and all multipoles (black curve). (e) and (f) show total normalized electric field intensity at the two points indicated in (a), respectively.}
\label{fig2}
\end{figure*}

The perpendicular scattering of plane waves by an ensemble of parallel nanowires can be calculated analytically through the multiple scattering method~\cite{BEVER_Appl.Opt._multiple_1992,FELBACQ_J.Opt.Soc.Am.AJOSAA_scattering_1994}. For the $j^{th}$ wire (centred at $\mathbf{r}_j$ on the transverse plane perpendicular to the longitudinal cylindrical axis) within such a scattering cluster of total wire number of $N$, its scattered fields outside the particles can be expanded into a series of cylindrical harmonics, and its scattering coefficients $a_{jm}^{s,p}$ are related to the single-wire scattering coefficients $a_m^{s,p}$ through:
\begin{equation}
\label{multiple_scattering}
a_{jm}^{s,p}  + a_{jm}^{s,p} \sum\limits_{q \ne j}^{q = 1:N} {\sum\limits_{l =  - \infty }^\infty  {\Omega_{jm,ql} a_m^{s,p} } }  = e^{-im\phi+\mathbf{k}\cdot \mathbf{r}_{j} } a_m^{s,p}.
\end{equation}
Here $\Omega_{jm,ql}$ is the coupling matrix element that characterizes the coupling between the $m^{th}$ cylindrical harmonic of the $j^{th}$ wire and the $l^{th}$ cylindrical harmonic of the $q^{th}$ wire:
\begin{equation}
\label{coupling_matrix}
\Omega_{jm,ql}=i^{l - m} \mathbf{H}^{(1)}_{l- m} (k|\mathbf{r}_{qj}| )e^{i(m - l)\phi _{qj}},
\end{equation}
where $\mathbf{H}^{(1)}$ is the Hankel function of the first kind, $\mathbf{r}_{qj}=\mathbf{r}_{j}-\mathbf{r}_{q}$, and $\phi _{qj}$ is the polar angle of $\mathbf{r}_{qj}$.  All sorts of couplings, including couplings between electric resonances, couplings between magnetic resonances, and cross magnetoelectric couplings between electric and magnetic resonances, have been embedded into $\Omega_{jm,ql}$ (\textit{e.g.}, $\Omega_{j0,q2}$ and $\Omega_{j2,q0}$ characterize the magnetoelectric coupling between MD and EQ for $p$-polarized waves). In this work we adopt the $e^{i(kr-\omega t)}$ notation for the electromagnetic waves. Through solving Eq.~(\ref{multiple_scattering}), the full sets of scattering coefficients of all nanowires $a_{jm}^{s,p}$ can be obtained, and then the scattering matrix is:
\begin{equation}
\label{scattering_matrix}
T^{s,p} (\alpha ) = \sum\limits_{j = 1}^N {\sum\limits_{m =  - \infty }^\infty  {e^{-i(m\alpha  + \mathbf{k_{\alpha}}\cdot\mathbf{r}_j)} } } a_{jm}^{s,p},
\end{equation}
where $\alpha$ is the scattering angle and $\mathbf{k_{\alpha}}=(k_x,k_y)=(k\cos\alpha,k\sin\alpha)$. It is worth mentioning that in coupled nanowire systems usually $a_{jm}^{s,p}\neq a_{j-m}^{s,p}$~\cite{WU_NanoLett._experimental_2015}, and for specific calculations, we would truncate the cylindrical harmonic number after the results become convergent.  Then according to the optical theorem~\cite{Bohren1983_book}, the extinction efficiency for the whole system of the coupled nanowires is:
\begin{equation}
\label{Q_ext_multi}
Q_{\rm ext}^{s,p} = {2\over {kR}}\mathbf{Re}[T^{s,p} (\alpha=0)],
\end{equation}
where for convenience the extinction efficiency is simply defined as the total extinction cross section divided by the diameter of a single wire ($2R$).  In a similar way, the fields inside each wire can be also expanded into a set of cylindrical harmonics with expansion coefficients $b_{jm}^{s,p}$, which can be obtained through matching the fields at the boundaries of the wires. With $a_{jm}^{s,p}$ and $b_{jm}^{s,p}$ at hand, both the far-field scattering properties and near-field distributions can be calculated analytically.

\section{Scattering properties of coupled nanowires with $p$-polarized waves}

As a first step we study the scattering of $p$-polarized incident waves by a quadrumer of $d=210$~nm and $\phi=0$, and show the total extinction efficiency spectra by solid curves in Fig.~\ref{fig2}(a). Two invisibility points with negligible scattering are indicated by points $\mathbf{A}$ and $\mathbf{B}$: $\lambda_A=425$~nm and $\lambda_B=787$~nm. We further calculate the scattering of such a quadrumer through the dipole approximation [considering only the ED and MD excitations of each nanowire and neglecting the higher order resonances by setting $a_{m}^{p}=0$ when $|m|\geq 2$ in Eq.~(\ref{multiple_scattering})]. The results are shown in Fig.~\ref{fig2}(a) the dashed curve, which do not agree with the full multipolar calculations (solid curve). As a result, in such a scattering cluster the contributions from higher order multipoles can not be simply neglected, which is actually induced by strong near-field coupling between close nanowires.

\subsection{Mechanisms of trivial and nontrivial invisibilities}
To reveal the different mechanisms associated with the two invisibility points identified above in Fig.~\ref{fig2}(a), we neglect the magnetoelectric cross couplings (the coupling between electric and magnetic multipoles) and conduct three approximations to further calculate the extinction efficiencies: (i) consider the contributions from MD only of each nanowire by setting $a_{m}^{p}=0$ when $m\neq0$ in Eq.~(\ref{multiple_scattering}); (ii) consider the contributions from electric multipoles of each nanowire only by setting $a_{m}^{p}=0$ when $m=0$ in Eq.~(\ref{multiple_scattering}); (iii) consider the contributions from both MD and electric multipoles, but neglect the cross coupling between them (while the coupling between MDs and coupling between electric multipoles are still taken into account) by setting $\Omega_{jm,ql}=0$ when $m\neq l$ and $l\cdot m=0$ in Eq.~(\ref{coupling_matrix}). The results are summarized in Fig.~\ref{fig2}(b). For all three approximations, the invisibility at point $\mathbf{A}$ can be almost reproduced  while none of them can reproduce the invisibility at point $\mathbf{B}$. This confirms the trivial nature of invisibility at point $\mathbf{A}$, where none of the multipoles of each individual nanowire have been effectively excited. This is quite clear from Fig.~\ref{fig1}(b), since at $\lambda_A=425$~nm the single nanowire is almost invisible itself and the scattering from all multipoles are negligible ($a_{m}^{p}\approx0$), which leads to the trivial scattering invisibility of the whole system [$a_{jm}^{p}\approx0$ ($j=1:4$) according to Eq.~(\ref{multiple_scattering})]. In contrast, the existence of the $\mathbf{B}$-point invisibility requires not only the simultaneous effective excitation of MDs and electric multipoles [see Fig.~\ref{fig1}(c)], but also the mutual magnetoelectric interactions between them, as is clear from Figs.~\ref{fig2}(a) and (b).

In the discussions above, we implement the approximations without magnetoelectric coupling through manipulating Eq.~(\ref{multiple_scattering}) and Eq.~(\ref{coupling_matrix}). To further clarify the mechanism of the invisibility obtained, we then take into considerations all the multipolar excitations and also the magnetoelectric cross coupling between them, but separate the contributions from electric multipoles and MDs from all nanowires.  To be more specific, the contributions from all the MDs can be obtained through setting $a_{j,m}^{p}=0$ when $m\neq0$ in Eq.~(\ref{scattering_matrix}), and those from electric multipoles only through setting $a_{j,m}^{p}=0$ when $m=0$ in Eq.~(\ref{scattering_matrix}). The results are shown in Fig.~\ref{fig2}(c), where for a better comparison we re-plot the results of full multipolar calculations.  It is clear from comparing Figs.~\ref{fig2}(b) and (c) that: at the trivial invisibility point, the effect of magnetoelectric coupling is negligible; while at the nontrivial invisibility point, the significant magnetoelectric interactions render the contributions from MDs and electric multipoles out of phase, making the particle cluster effectively invisible.

We emphasize that in passive scattering systems (without gain media), energy conservation requires that the scattering cross-section associated with each eigenmode of the whole system can not be negative, as such an eigenmode can be independently excited. What we have shown in Fig. ~\ref{fig2}(c) includes the separate contributions from the MDs and all electric multipoles, which nevertheless are not independent eigenmodes of the whole system, as the magnetoelectric coupling between them clearly exists [comparing Figs.~\ref{fig2}(b) and (c)]. Mathematically the negative and positive values of the contributions to extinction cross sections from MDs and electric multipoles shown in Fig. ~\ref{fig2}(c) can be understood from Eqs.~(\ref{scattering_matrix})-(\ref{Q_ext_multi}). Physically it means nothing more than that the two sets of responses interfere destructively, the scatterings of which cancel each other in the far filed and thus result in the nontrivial invisibility. Actually such an analysis relying on responses that are not independent are widely employed in structures that support electromagnetically induced transparency~\cite{fleischhauer_electromagnetically_2005,YANG_NatCommun_alldielectric_2014,Zhang2008_PRL}, Fano resonances~\cite{lukyanchuk_fano_2010-1}, or anapole modes ~\cite{Basharin2014_arXiv,miroshnichenko2014seeing,liu_toroidal_2015,Liu2015_OL_invisible,LIU_ArXivPrepr.ArXiv160901099_multipolar_2016,PAPASIMAKIS_NatMater_electromagnetic_2016}, where the simplest model of two-mode interactions plays a fundamental role.

It is worth mentioning that at point $\mathbf{B}$ in Fig.~\ref{fig2}(c), though both the contributions are zero, it does not mean that $a_{jm}^{p}\approx0$ ($j=1:4$) (which nevertheless is exactly the case at point $\mathbf{A})$. This is clear in Fig.~\ref{fig2}(d) where we show the sum of the scattering coefficients intensity (corresponding to a simple sum of the overall scattering intensities of all nanowires~\cite{Bohren1983_book}): $\Lambda^{p} = \sum\nolimits_{j = 1}^4 {\sum\nolimits_{m = - \infty }^{\infty}} {|a_{jm}^{p}|^{2}}$. The partial contributions from MDs ($\sum\nolimits_{j = 1}^4{|a_{j0}^{p}|^{2}}$) and electric multipoles ($ \sum\nolimits_{j = 1}^4 {\sum\nolimits_{m \neq 0}} {|a_{jm}^{p}|^{2}}$) are also shown. As is clearly shown, at point $\mathbf{B}$ there are significant multipolar excitations, which however are negligible at point $\mathbf{A}$. The different natures of the two points can be further justified by calculating their corresponding near-field distributions, as is shown in Figs.~\ref{fig2}(e) and (f).  Here we show the total electric field intensity (normalized by the electric field of the incident wave) at the two invisibility points discussed, and it is clear that at the nontrivial point $\mathbf{B}$ there is significant free-space field enhancement within the gaps between the nanowires, which originate from effective multipolar excitations within each nanowire, and thus is much more significant than the more or less negligible enhancement at the trivial point $\mathbf{A}$.

\begin{figure*}
\centerline{\includegraphics[width=18cm]{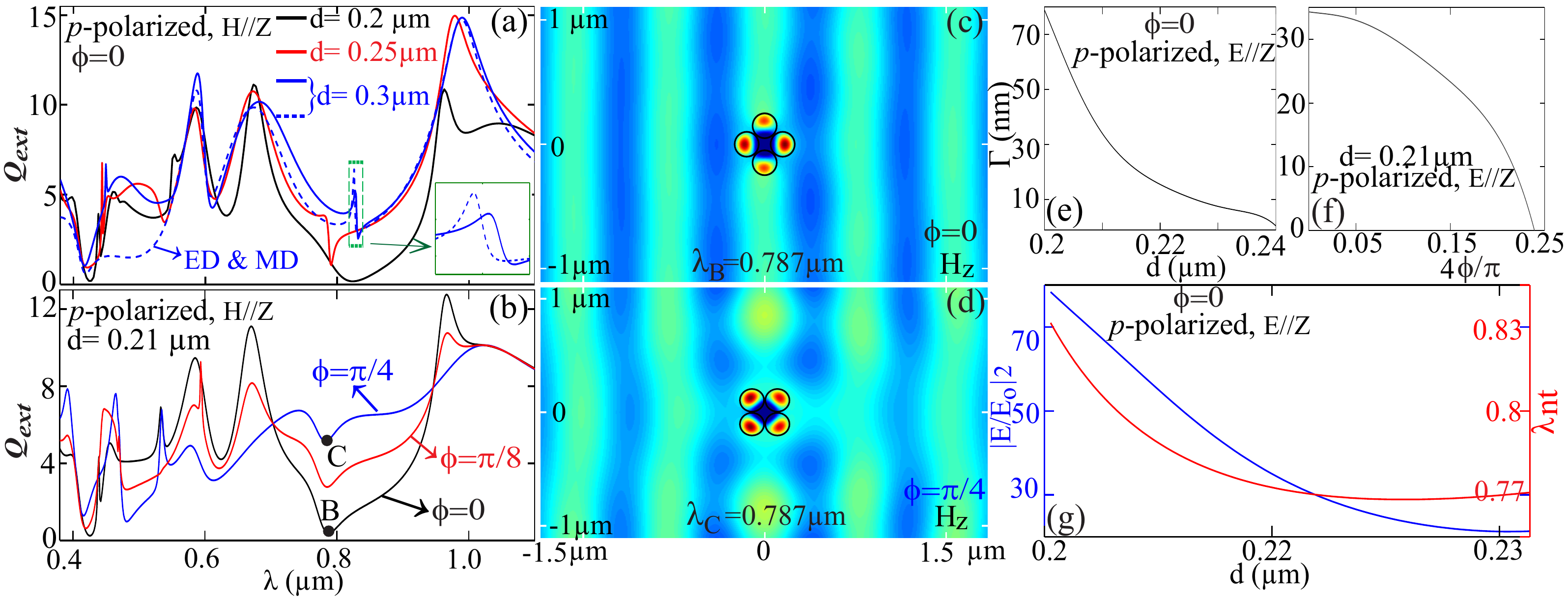}}\caption{Extinction efficiency spectra with $p$-polarized incident waves for the quadrumer of different inter-particle distances in (a) with fixed $\phi=0$: $d=200,~250,~300$~nm, and of different propagation directions in (b) with fixed $d=210$~nm: $\phi=0,~\pi/8,~\pi/4$. The results obtained with the dipole approximation are shown also in (a) for $d=300$~nm (dashed blue curve), and the marked area is shown enlarged as the inset. Two minimum scattering points at the same spectral position of $\lambda=787$~nm are marked in (b) by points $\mathbf{B}$ ($\phi_{B}=0$) and $\mathbf{C}$ ($\phi_C=\pi/4$), and the magnetic field distributions ($\mathbf{H}_z$) at those two points are shown in (c) and (d), respectively. (e) and (f) show respectively the dependence of $\Gamma$ on $d$ (with $\phi=0$) and on $\phi$ (with $d=210$~nm), respectively at the nontrivial invisibility region. (g) Spectral positions ($\lambda_{\rm nt}$) of the nontrivial invisibility dips (red curve) and the field  enhancement (normalized electric field intensity at the center between adjacent particles) at those dips (blue curve) with different $d$.}
\label{fig3}
\end{figure*}

\subsection{Dependence of the scattering properties on inter-particle distance $d$ and incident angle $\phi$}
As a next step, we investigate the dependence of the invisibility obtained on the inter-particle distance $d$ and the propagation direction characterized by $\phi$. The results are summarized in Figs.~\ref{fig3}(a) and (b).  As is expected, the trivial invisibility obtained is slightly dependent on $d$ and $\phi$, since at this point each nanowire is very weakly excited and thus the change of $d$ and $\phi$ will not affect significantly the multipolar coupling or the scattering properties. In contrast, the nontrivial invisibility obtained is significantly dependent on both $d$ and $\phi$, and at some regimes the invisibility can be eliminated. This is because the nontrivial invisibility demonstrated above relies on strong excitations and couplings between the electric and magnetic multipoles, and different $d$ or $\phi$ lead to quite different couplings, resulting in totally distinct scattering properties. For the case of relatively large inter-particle distance of $d=300$~nm, the extinction spectra from both full multipolar calculation and dipole approximation are shown in Fig.~\ref{fig3}(a). It is clear that even for such large inter-particle distance, the dipole approximation is still insufficient [see the inset of Fig.~\ref{fig3}(a)], which further proves the strong near-field coupling between the nanowires and can thus explain the multipolar interference origin and strong dependence of the nontrivial invisibility. At the same time, we select two points at the same spectral position ($\lambda=787$~nm) in Fig.~\ref{fig3}(b), and show the analytically calculated corresponding near-field distributions (in terms of $\mathbf{H}_z$) in Figs.~\ref{fig3}(c) and (d). As is shown, changing the propagation direction can switch the effectively invisible particle clusters [see Fig.~\ref{fig3}(c)] to be visible, where the incident wave  experiences significant perturbations [see Fig.~\ref{fig3}(d)]. This is also the case for different $d$, as is indicated in  Fig.~\ref{fig3}(a): the invisibility point is preserved for relatively small $d$ and exists even for touching nanowires of $d=2R=200$~nm; for larger $d$, the spectra will evolve into a typical Fano profile~\cite{Miroshnichenko2012_NL6459,HOPKINS_ACSPhotonics_interplay_2015,YAN_ACSNano_directional_2015}, while at the same time the invisibility will gradually disappear.

\subsection{Dependence of the nontrivial invisibility spectral width on inter-particle distance $d$ and incident angle $\phi$}
To more accurately characterize the dependence of the nontrivial invisibility on $d$ and $\phi$, we find the spectral regime when the particles are effectively non-trivially invisible and define the spectral width of such region as $\Gamma$. Though there is no universal clear boundary between being visible or invisible, here for convenience we view the quadrumer as invisible when the scattering cross section is smaller than the diameter of a single nanowire ($Q_{\rm ext}\leq 1$), which at the same time is also far smaller than the wavelength. This means that $\Gamma=\lambda_l-\lambda_s$, where $\lambda_l$ ($\lambda_s$) is the larger (smaller) wavelength at which $Q_{\rm ext}=1$. The dependence of $\Gamma$ on $d$ (with $\phi=0$) and on $\phi$ (with $d=210$~nm) at the nontrivial invisibility region is shown in Figs.~\ref{fig3}(e) and (f), and it is clear that larger $d$ and/or large $\phi$ will shrink the invisibility spectral width.  For closed packed nanowires with small $\phi$, there is an effectively wide spectral regime where we can obtain simultaneous invisibility and free-space field enhancement. This originates from that the contributions from MDs and electric multipoles are out of phase in a wider spectral region centered at the nontrivial invisibility point [see Fig.~\ref{fig2}(c)], and it might be employed for non-invasive signal extraction. We note that what we obtain here is very different from that achieved in Ref.~\cite{YAN_ACSNano_directional_2015}, where despite the fact that they employ an extended metasurface, the invisibility and free-space field-enhancement coexist only within a much narrower spectral region in their work.

We also characterize the dependence of the field enhancement on $d$ and summary the results in Fig.~\ref{fig3}(g) (with $\phi=0$). According to Fig.~\ref{fig3}(a), for different inter-particle distances, the nontrivial invisibility regions would spectrally shift and we use $\lambda_{\rm nt}$ to specifically denote the spectral position of the the nontrivial invisibility  dip [such as point $\textbf{B}$ in Fig.~\ref{fig2}(a) and Fig.~\ref{fig3}(b)]. Figure~\ref{fig3}(g) shows the dependence of both $\lambda_{\rm nt}$  and the free-space field intensity enhancement (when $\lambda=\lambda_{\rm nt}$) at the points located at the middle of the lines connecting the centers of the adjacent particles [$(x,y)=(\pm\frac{d}{2\sqrt{2}},\pm\frac{d}{2\sqrt{2}})$; see the maximum field points in Fig.~\ref{fig2}(f)].

\begin{figure}
\centerline{\includegraphics[width=7cm]{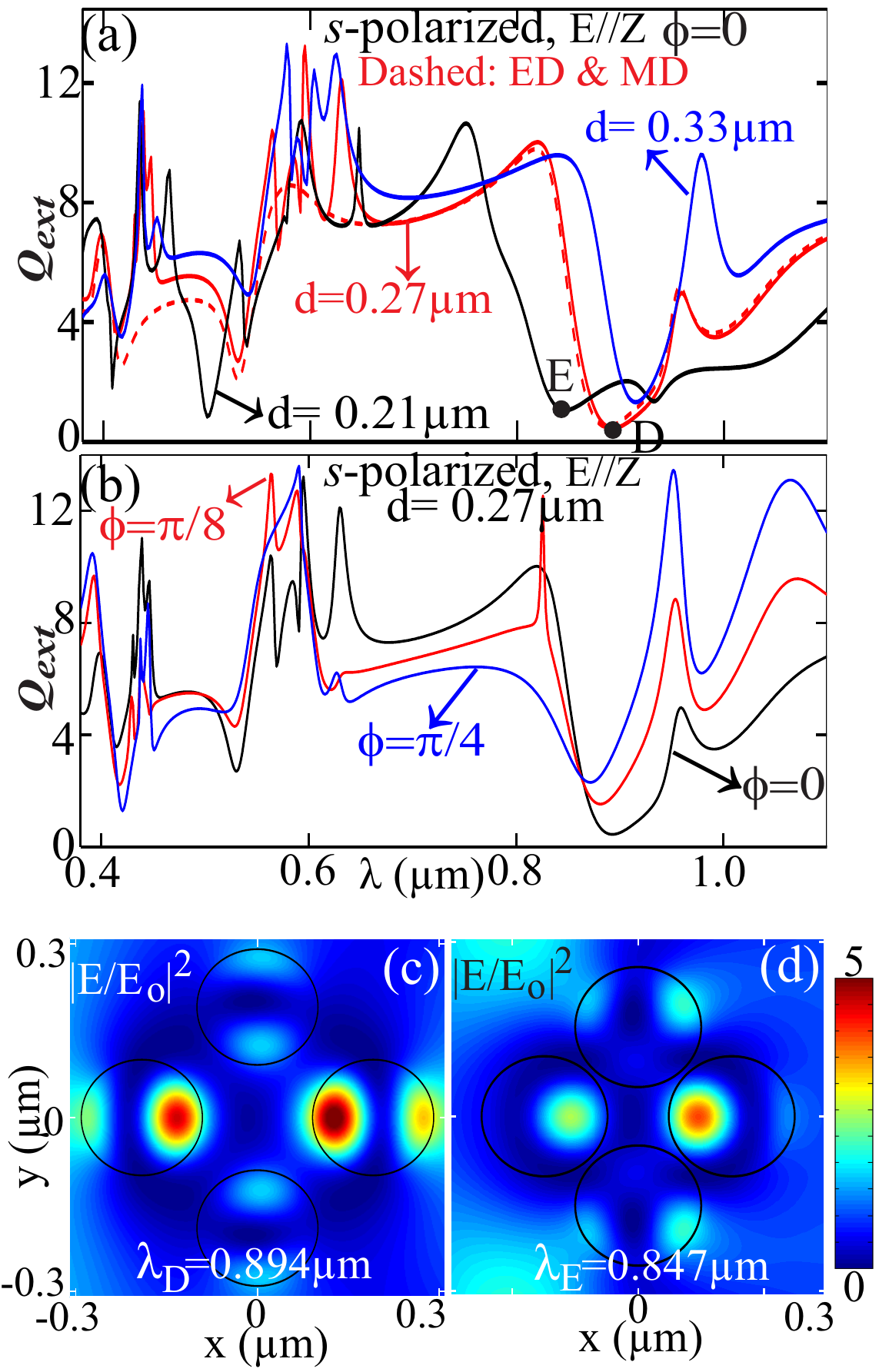}}\caption{Extinction efficiency spectra with $s$-polarized incident waves for the quadrumer of different inter-particle distances in (a) with fixed $\phi=0$: $d=210,~270,~330$~nm, and of different propagation directions in (b) with fixed $d=270$~nm: $\phi=0,~\pi/8,~\pi/4$. The results obtained under dipole approximation are shown also in (a) for $d=270$~nm (dashed red curve). Two minimum scattering points are indicated in (a) by points $\mathbf{D}$ ($\lambda_D=894$~nm) and $\mathbf{E}$ ($\lambda_E=847$~nm), and the total normalized electric field intensity at those two points are shown in (c) and (d).}
\label{fig4}
\end{figure}

\section{Scattering properties of coupled nanowires with $s$-polarized waves}

At the end we investigate the quadrumer scattering of $s$-polarized incident waves. Figure~\ref{fig4}(a) shows the extinction efficiency spectra with fixed $\phi=0$ but of different inter-particle particle distances. For $d=270$~nm, an effective invisibility point can be achieved and is indicated by point $\mathbf{D}$ ($\lambda_D=894$~nm), which in a similar way is induced by the coexistence and coupling between electric and magnetic resonances [see point $\mathbf{B}$ in Fig.~\ref{fig2}(a) and (b)]. As is demonstrated, such invisibility is dependent on both $d$ [Fig.~\ref{fig4}(a)] and $\phi$ [Fig.~\ref{fig4}(b)]. Compared to the case of $p$-polarization shown in Figs.~\ref{fig3}(a) and (b), the dependence here is relatively much less significant, due to the fact that with $s$-polarization the near-field interactions are much weaker.  This is also clear in Fig.~\ref{fig4}(a) for the case of $d=270$~nm, where we show both the results obtained by full multipolar calculation and by the dipole approximation, which agree quite well with each other within quite a broad spectral regime. It means that the contributions from higher order multipoles are negligible, which suggests weak near-field interactions.  For the case of $p$-polarization however, as is shown in Fig.~\ref{fig3}(a), for an even larger inter-particle distance $d=300$~nm, the dipole approximation is still insufficient, indicating stronger near-field interactions.

To further clarify this point, we select two points $\mathbf{D}$ and $\mathbf{E}$ ($\lambda_E=847$~nm) in Fig.~\ref{fig4}(a) and show the analytically calculated total electric field intensity in Figs.~\ref{fig4}(c) and (d), respectively.  In sharp contrast to the case of $p$-polarization where most of the fields are trapped within the gaps between the particles [see Figs.~\ref{fig2}(e) and (f)], for $s$-polarization most of the fields are confined within the particles, which results in much weaker near-field inter-particle coupling and also weaker dependence of both extinction spectra and near-field enhancements on $d$ [see Figs.~\ref{fig4}(c) and (d)] and $\phi$. This is quite comprehensible: with $p$-polarized incident waves, the electric fields are distributed on-plane ($x-y$ plane), which are enhanced close to the nanowire boundary at the outside sides due to large index  mismatch; with $s$-polarization however, there is only $\mathbf{E}_z$ component, which is parallel to and continuous across the boundary, and thus not enhanced outside of the particles. As a result, at the invisibility points of $\mathbf{B}$ and $\mathbf{D}$, the near-field enhancement is much more significant for $p$-polarization [see Fig.~\ref{fig2}(f)] than for $s$-polarization [Fig.~\ref{fig4}(c)]. In a word, for $s$-polarization though we obtain multipolar interference induced invisibility with each nanowire effectively excited, there is no simultaneous significant free-space field enhancement. This is a special case between the trivial and nontrivial invisibilities discussed for the $p$-polarization.

\section{Conclusions and outlook}
To summarize, here in this work we demonstrate both trivial and nontrivial invisibilities in all-dielectric quadrumers consisting of coupled high-index parallel nanowires.  The trivial invisibility relies on the effectively null excitation of each individual nanoparticle, while the nontrivial invisibility together with free-space field enhancement originate from not only the significant excitations of both electric and magnetic multipoles of each nanowire, but also from their significant mutual magnetoelectric cross-interactions. We further show that the invisibility obtained is sensitive to both the polarizations and shining angles of the incident waves. The functioning mechanism behind those phenomena we demonstrate can hopefully play a significant role in variously applications and fundamental researches related to scattering nanoparticle systems.

We note that here in this work we confine our study to symmetric quadrumers consisting of identical parallel nanowires. The principles can certainly be extended to particles of other shapes (such as spheres and many other experimentally more  feasible irregular structures) and of other distributions (being it random or periodic), and to other spectral regimes considering that photonic dielectric structures are fully scalable. It would be interesting to introduce such high index particles into novel scattering systems and to investigate the interplay of multipolar interactions with various effects including photonic bandgaps and spectral gaps, random lasing, Anderson localizations, $\mathcal{P}\mathcal{T}$-symmetric electromagnetic scattering, and optical topological effects, which can potentially incubate many fresh fundamental ideas and new applications.

We thank Ben Hopkins and Yuri S. Kivshar for useful discussions and acknowledge a financial support from the National Natural Science Foundation of China (Grant number: $11404403$), the Australian Research Council and the Outstanding Young Researcher Programme of the National University of Defense Technology.


\end{document}